\documentclass[12pt]{article}

\catcode`\@=11
\@addtoreset{equation}{section}

\global\arraycolsep=2pt
\usepackage{amsbsy,amssymb,latexsym,amsfonts,amsmath}
\usepackage{graphicx,color}

\allowdisplaybreaks

\begin{document}
\begin{flushright}
\parbox{4.2cm}
{UCB-PTH-09/36, IPMU09-0165}
\end{flushright}

\vspace*{0.7cm}

\begin{center}
{\Large \bf Anisotropic scale invariant cosmology}
\vspace*{2.0cm}\\
{Yu Nakayama}
\end{center}
\vspace*{-0.2cm}
\begin{center}
{\it Berkeley Center for Theoretical Physics, \\ 
University of California, Berkeley, CA 94720, USA \\

and  \\

Institute for the Physics and Mathematics of the Universe, \\
University of Tokyo, Kashiwa, Chiba 277-8582, Japan
}
\vspace{3.8cm}
\end{center}

\begin{abstract} 
We study a possibility of anisotropic scale invariant cosmology. 
It is shown that within the conventional Einstein gravity, the violation of the null energy condition is necessary.
We construct an example based on a ghost condensation model that violates the null energy condition. The cosmological solution necessarily contains  
at least one contracting spatial direction as in the Kasner solution. Our cosmology is conjectured to be dual to, if any, a non-unitary anisotropic scale invariant Euclidean field theory. We investigate simple correlation functions of the dual theory by using the holographic computation. After compactification of the contracting direction, our setup may yield a dual field theory description of the winding tachyon condensation that might solve the singularity of big bang/crunch of the universe.

\end{abstract}

\thispagestyle{empty} 

\setcounter{page}{0}

\newpage

\section{Introduction}
Gauge/gravity correspondence \cite{Maldacena:1997re}, or more broadly holography \cite{hol1}\cite{hol2} is a key to understand non-perturbative features of quantum gravity. Cosmology is a natural arena where we can apply the holographic technique to understand the time evolution of the universe, the landscape program, and the initial singularity of the big bang. While the foundation of the cosmological applications of gauge/gravity correspondence is less established than the standard AdS/CFT correspondence, there are many ambitious attempts including \cite{Strominger:2001pn}\cite{Strominger:2001gp}\cite{Freivogel:2006xu}\cite{Sekino:2009kv}.

On the other hand, anisotropic scale invariant gravity solutions have attracted  a lot of attentions these days, primarily focusing on their applications to condensed matter physics \cite{Kachru:2008yh}. The solution has a non-relativistic dispersion relation and it has been argued that it may give a dual description of the strongly coupled limit of the Lifshitz-like scale invariant field theories \cite{Lif}. Furthermore, an alternative proposal for ultra-violet completion of general relativity has been pushed forward based on the anisotropic non-relativistic gravity action \cite{Horava:2008ih}\cite{Horava:2009uw}\cite{Horava:2009if}. The breaking of the Poincar\'e invariance has played a significant role in such examples.
The Poincar\'e invariance, which we believe to be true in the low-energy limit of our daily lives may not be the fundamental principle of physics. General relativity does not require that the solution should be Poincar\'e invariant, and neither does the string theory. It is an emergent symmetry. Besides, the cosmological time evolution explicitly breaks the Poincar\'e invariance.

In this paper, we would like to combine the idea of the holographic cosmology and the anisotropic scale invariance within the conventional general relativity. We examine a possibility of anisotropic scale invariant cosmology. The cosmological solution has a holographic interpretation of the Euclidean anisotropic scale invariant field theory that might be realized in a condensed matter system.

The inspection of the Einstein equation tells us that the anisotropic scale invariant cosmology is only possible within the conventional Einstein gravity when the null energy condition is violated in the matter sector. While there could be many possible ways to introduce rather exotic matters to avoid the constraint, we show one particular example based on the ghost scalar action with time-like condensation \cite{ArkaniHamed:2003uy}. The resulting anisotropic cosmology always contains at least one contracting spacial direction much like the Kasner solution of the vacuum Einstein equation.

The dual field theory is most presumably non-unitary as indicated by the holographic correlation functions. However, it turns out that this non-unitarity is exactly what is needed to tame the long range growing correlation functions of the dual field theory. Typically, a contracting spatial direction in the holographic cosmology would lead to an instability of the dual field theory, but the pure imaginary scaling dimension makes it behave oscillating rather than growing.

In Euclidean field theories, the unitarity, or more precisely the reflection positivity, is not the holy grail at all. There are many physical examples where the unitarity is violated. The conventional AdS/CFT may not be suitable to provide a dual gravity description of such systems, and the cosmological setup can be useful for this purpose. Unfortunately, the consistency of such cosmological models beyond the gravity approximation is a delicate issue, and we may eventually need an embedding in the string theory or something ultra-violet completed. Our approach is rather bottom up, and ultra-violet completion of the system will be studied elsewhere in the future.

\section{Anisotropic scale invariant cosmology}
Let us study the Lifshitz-like anisotropic scale invariant cosmology in (1+3) dimension. Generalization to higher (lower) dimension will be obvious. Our scale invariant cosmological ansatz for the metric is given by
\begin{align}
ds^2 = -\frac{dt^2}{t^2} + \frac{dx^2}{t^{2a}} + \frac{dy^2}{t^{2b}} + \frac{dz^2}{t^{2c}} \ . \label{anis}
\end{align}
The metric is invariant under the anisotropic scaling $t \to \lambda t$, $ x \to \lambda^a x$, $y \to \lambda^b y$, and $z \to \lambda^c z$. The ansatz is consistent with the translational invariance in $(x,y,z)$ as well as the parity invariance $x^i \to -x^i$. 
By a coordinate transformation, one can always choose one of the three dynamical scaling exponents (say $a$) to be one. A special choice $a=b=c$ corresponds to de-Sitter space.

We first show that except for this particular de-Sitter case, which is isotropic, the anisotropic scale invariant cosmology \eqref{anis} is only possible  within the conventional Einstein gravity when the null energy condition is violated. To see this, we compute the Einstein tensor for the metric \eqref{anis} as:
\begin{align}
G_{tt} &= \frac{bc + ab + ac}{t^2} \ , \ \ \ G_{xx} = -\frac{b^2+bc+c^2}{t^{2a}} \cr
G_{yy} &= -\frac{c^2 + ca+a^2}{t^{2b}} \ , \ \ G_{zz} = -\frac{a^2 + ab + b^2}{t^{2c}} \ . 
\end{align}
Now, the null energy condition demands $G_{\mu\nu}k^{\mu}k^{\nu} \ge 0 $ for any null vector $k^{\mu}$. By taking $k^{\mu} = (\sqrt{3}t, t^{a}, t^{b},t^{c})$, we obtain 
\begin{align}
-(a-b)^2 - (b-c)^2 - (c-a)^2 \ge 0  \ ,
\end{align}
which is only possible when $a=b=c$. Thus, except for the special case of de-Sitter cosmology, the anisotropic scale invariant cosmology is inconsistent with the null energy condition.\footnote{A related Bianchi I cosmology was studied in \cite{Aref'eva:2009vf}\cite{Aref'eva:2009xr}, where the violation of the null energy condition and its (in)stability were investigated. We would like to thank I.~Aref'eva for the correspondence.}

In order to realize the anisotropic scale invariant cosmology within the Einstein gravity, therefore,  it is necessary to break the null energy condition. While there could be many ways to do this, here, we would like to investigate the possibly by using the ghost matter \cite{ArkaniHamed:2003uy}:
\begin{align}
 S = \int d^4x \sqrt{-g} F(\partial^\mu \phi \partial_\mu \phi) \ ,
\end{align}
where non-trivial $F(X)$ with $X=\partial^\mu \phi \partial_\mu \phi$ introduces generic higher derivative interaction consistent with the shift symmetry $\phi(x) \to \phi(x) + \Lambda$. 
The ansatz for scalar field $\phi$ for the scale invariant cosmology without breaking the translational invariance is 
\begin{align}
 \phi = p \log t \ . \label{fig}
\end{align}
As in \cite{Nakayama:2009qu}, we have to gauge the constant shift of the scalar field $\phi(x) \to \phi(x) + \Lambda$ so that the scaling transformation is a symmetry of the ansatz. 

The equation of motion for $\phi$ is solved either by $F'(-p^2)=0$ or $a + b + c = 0$. We focus on the latter case $a+b+c=0$.\footnote{The former solution leads to $a=b=c$, and hence the de-Sitter space, which is identical to the scale invariant but non-conformal scalar field configuration studied in \cite{Nakayama:2009qu}.} The field configuration \eqref{fig} now supplies the additional negative energy in the energy momentum tensor in addition to the cosmological constant when $F'(-p^2) <0$: $T_{\mu\nu} = -\tilde{\Lambda} g_{\mu\nu} + \text{diag}(F'(-p^2),0,0,0)$, where $\tilde{\Lambda}$ is the effective cosmological constant we have introduced to solve the Einstein equation. For this particular form of the energy momentum tensor, the Einstein equation is solved by demanding $ a + b + c =0$, which coincides with the condition that the scalar field ansatz \eqref{fig} solves the ghost equation of motion.

The fluctuation around the ghost condensation \eqref{fig} has negative energy and possibly non-unitary spectrum depending on the boundary condition we impose. A consistency of such configuration has been doubted in \cite{Nakayama:2009qu}: in particular when $a=b=c$, the configuration with assumed unitarity in the boundary theory is inconsistent with Polchinski's theorem that states the unitary Lorentz and scale invariant theory must be conformally invariant \cite{Polchinski:1987dy}\cite{Zamolodchikov:1986gt}\cite{Nakayama:2009fe}. As we will see in the next section, the boundary theory for our configuration is most presumably not unitary, and it is not entirely clear why the gravity theory should be so, either. On the other hand, we might be able to come up with a better matter sector that is consistent with unitarity while violating the null energy condition to realize the anisotropic scale invariant cosmology.\footnote{For instance, the orientifold in string theory can violate the null energy condition, so the violation of the null energy condition itself may not be inconsistent with the consistency of quantum theories of gravity.} In the computation of the correlation functions, therefore, we only focus on the universal geometric background and will not discuss the ghost matter sector to be on the optimistic side.

The condition $a+b+c=0$ means that at least one spatial direction is contracting. One may compactify the contracting direction so that the visible universe is expanding while the internal space is contracting.\footnote{The continuous scale invariance will be broken by the compactification. On the other hand, we may expect winding tachyon condensation that might cure the big bang/crunch singularity of our cosmology \cite{McGreevy:2005ci}\cite{Nakayama:2006gt}. The discussions in the following sections might give a holographic dual descriptino of such a scenario.
} In particular, when $a=b = -1/2 c$, the anisotropic cosmology admits additional rotational invariance, and the symmetry algebra is Wick rotated version of the Lifshitz-like scale invariant theory whose gravity dual was first proposed in \cite{Kachru:2008yh}. A difference here is the ``dynamical critical exponent" is now negative. In their setup, the energy condition has yielded a constraint that the dynamical critical exponent is greater than one. Their constraint may have a physical meaning due to the finiteness of the speed of light. Thanks to the Euclidean signature of our dual theory, such constraints cannot appear here.

Our metric resembles the Kasner solution of the vacuum Einstein equation:
\begin{align}
ds^2 = -d\tau^2 + \tau^{2a} dx^2 + \tau^{2b} dy^2 + \tau^{2c} dz^2 \ ,
\end{align}
where $a + b + c = 1$ and $a^2 + b^2 + c^2 = 1$. Note that one of the exponents is always negative similarly to ours. It is instructive to rewrite our metric by introducing $t = e^{\tau}$ as
\begin{align}
ds^2 = -d\tau^2 + e^{2a\tau} dx^2 + e^{2b \tau} dy^2 + e^{2c \tau} dz^2 \ .
\end{align}
One may regard it as an exponentially expanding/contracting version of the Kasner universe. 

It is well-known that by further relaxing the condition of ``flatness", the alternating feature of the Kasner regime appears near the singularity at the beginning of the universe. It would be interesting to study a similar situation in our anisotropic scale invariant cosmology. 

\section{Correlation functions from holography}
The gauge/gravity correspondence is a non-perturbative way to understand the dual field theories.  Alternatively, one may understand the nature of quantum gravity from the dual field theories. 
Originally, it was proposed in the negatively curved space like AdS space \cite{Gubser:1998bc}\cite{Witten:1998qj}, while some attempts have been done to generalize it in the cosmological setup \cite{Strominger:2001pn}. We would like to use the holographic technique to compute correlation functions to understand the nature of the dual field theory of our anisotropic scale invariant cosmology, if any. At the same time, the holographic computation further reveals some peculiar features of the anisotropic scale invariant cosmology.

We introduce a conventional scalar field $\varphi$ with mass $m$ that is minimally coupled to the Einstein gravity:
\begin{align}
S = \int d^4 x \sqrt{-g}\left( \partial_\mu \varphi \partial^\mu \varphi - m^2 \varphi^2 \right) \ 
\end{align}
to compute the holographic correlation functions among the operator $O$ associated with the scalar $\varphi$.

The equation of motion for the scalar is given by
\begin{align}
t^2 \partial_t^2 \varphi + t \partial_t \varphi - t^{2a} \partial_x^2 \varphi - t^{2b}\partial_y^2 \varphi - t^{2c}\partial_z^2 \varphi + m^2 \varphi = 0 \ . \label{eoms}
\end{align}
From the translational invariance, it is convenient to go to the momentum space $\varphi = \tilde{\varphi}(t,k) e^{ik_x x + ik_y y + ik_z z}$. A standard holographic recipe to compute the two-point function gives (here $\epsilon$ is an IR cutoff)
\begin{align}
\langle O(k)O(p) \rangle = \delta(k+p) \left[\tilde{G}(t,-k)\sqrt{-g}g^{tt}\partial_t \tilde{G}(t,k)\right]_{\epsilon}^{\infty}
\end{align}
by using the bulk boundary propagator $\tilde{G}$ associated with \eqref{eoms} so that 
\begin{align}
\tilde{\varphi}(t,k) = \tilde{G}(t,k) \tilde{\varphi}(0,k) \ . 
\end{align} 

We could not  find an analytic expression for the most general solution, so we first discuss the asymptotic form of the solution. 
Without loosing generality, we assume $c < 0 \le b \le a$: if two of the exponents are negative, one can perform the coordinate transformation $t\to t^{-1}$ to retain the inequality. For $t \ll 1$, the solution for non-zero $k_x$ is given by the Bessel functions: 
\begin{align}
\tilde{\varphi} = J_{\pm \frac{im}{a}} \left(\frac{k_xt^a}{a}\right) = \left(\frac{k_xt^a}{a}\right)^{\frac{\pm im}{a}} \left[\frac{2^{\mp \frac{im}{a}}}{\Gamma(1\pm \frac{im}{a})} - \frac{2^{-2\mp \frac{im}{a}}\left(\frac{k_x t^a}{a}\right)^2}{\Gamma(2\pm \frac{im}{a})} + \cdots \right] \ .
\end{align}
On the other hand, for $t\gg 1$, the solution for non-zero $k_z$ is given by
\begin{align}
\tilde{\varphi} = J_{\pm \frac{im}{c}} \left(\frac{k_z t^c}{c}\right) = \left(\frac{k_xt^c}{c}\right)^{\frac{\pm im}{c}} \left[\frac{2^{\mp \frac{im}{c}}}{\Gamma(1\pm \frac{im}{c})} - \frac{2^{-2\mp \frac{im}{c}}\left(\frac{k_z t^c}{c}\right)^2}{\Gamma(2\pm \frac{im}{c})} + \cdots \right] \ . \ .\end{align}
Note that unlike the Euclidean case, there is no compelling principle to choose a particular linear combination of the solution of the equation of motion: the choice will be reflected in the ambiguity to choose propagators (and vacuum) in the dual field theory.\footnote{Recently, more detailed studies on the anisotropic conformal boundary conditions have been presented in \cite{Horava:2009vy}.} For instance, in the conventional (Euclidean) AdS/CFT setup, it is customary to choose a particular linear combination given by the modified Bessel function:
\begin{align}
\tilde{\varphi} = K_{\frac{im}{a}}  \left(\frac{k_xt^a}{a}\right)
\end{align}
near $t \ll 1$. That would correspond to choosing the Feynman propagator in the conventional AdS/CFT correspondence.

The scaling dimension of the operator is given by 
\begin{align}
\Delta (O) = \pm im \label{sd}
\end{align}
The imaginary scaling dimension for real $m$ suggests the dual field theory is not unitary. Another possibility is that the theory is unitary, but the spectrum probed by the holography does not have corresponding states as in imaginary conformal dimension operators in Liouville theory (e.g. \cite{Nakayama:2004vk} for a review). In this case, the analytic continuation of the conformal dimension is motivated, and we will come back to these points later.

Let us study some simple correlation functions. When only one of the momentum is excited, they are given by
\begin{align}
\langle O(k_x) O(p_x) \rangle &= \delta(k_x + p_x) \frac{1}{k_x^{\pm \frac {im}{a}}} \ , \cr  
\langle O(k_y) O(p_y) \rangle &= \delta(k_y + p_y) \frac{1}{k_y^{\pm \frac {im}{b}}} \ , \cr 
 \langle O(k_z) O(p_z) \rangle &= \delta(k_z + p_z) \frac{1}{k_z^{\pm \frac {im}{c}}} \ . \label{sic}
\end{align}
The sign of the scaling dimension in \eqref{sd} corresponds to the choice of propagators.


Another particular solvable case is $a=2b$. For $a = 2$, the explicit form of the solution for $k_z = 0$ is 
\begin{align}
\varphi &= c_1e^{-i\frac{k_xt^2}{2}} t^{1+im} U \left(i\frac{k_y^2}{4k_x} + \frac{1}{2} + \frac{im}{2}, 1+ im, ik_xt^2 \right)\cr
 &+ c_2 e^{-i\frac{k_xt^2}{2}} t^{1+im} L\left(-i\frac{k_y^2}{4k_x} - \frac{1}{2} - \frac{im}{2}, im, ik_xt^2 \right) \ 
\end{align}
by using confluent hypergeometric functions (see appendix A for details).
A suitable choice of the solution corresponds to the analytic continuation of the two-point function in the Lifshitz geometry studied in \cite{Kachru:2008yh} (see also \cite{Nakayama:2009ww}):
\begin{align}
\langle O(k_x,k_y) O(p_x,p_y) \rangle &= \delta(k_x+p_x)\delta(k_y+p_y) k_x^{-im} \frac{\Gamma\left(i\frac{k_y^2}{4k_x}+\frac{1}{2} -\frac{im}{2}\right)}{\Gamma\left(i\frac{k_y^2}{4k_x}+\frac{1}{2} +\frac{im}{2}\right)} \label{lift}
\end{align}
up to an overall normalization factor.

So far, all the operators corresponding to massive scalar had the pure imaginary conformal dimension. However, we could introduce a scalar field with the negative mass squared that would correspond to real conformal dimension. In dS/CFT \cite{Strominger:2001pn}, the corresponding statement would be to study ``unstable" scalar modes that would correspond to the real conformal dimension. Actually, the conventional recipe of the AdS/CFT correspondence to compute correlation functions can be best suited in this ``unstable" range of mass parameters because the distinction between the normalizable modes and non-normalizable modes are clearly displayed, and the ambiguity to choose the propagator is less apparent.

Yet another application of the ``tachyonic mode" here is the winding tachyon condensation studied in \cite{McGreevy:2005ci}\cite{Nakayama:2006gt}. Our prescription gives a holographic dual description of the winding tachyon condensation by identifying the scalar mode as the winding tachyon (up on T-duality).

In our case, \eqref{sic} can be analytically continued to
\begin{align}
\langle O(k_x) O(p_x) \rangle &= \delta(k_x + p_x) \frac{1}{k_x^{\frac {\tilde{m}}{a}}} \ , \cr 
\langle O(k_y) O(p_y) \rangle &= \delta(k_y + p_y) \frac{1}{k_y^{\frac {\tilde{m}}{b}}} \ , \cr 
 \langle O(k_z) O(p_z) \rangle &= \delta(k_z + p_z) \frac{1}{k_z^{\frac {\tilde{m}}{c}}} \ ,
\end{align}
where $\tilde{m} = \pm i m > 0$. In particular, the two-point function would be growing in large $z$ direction, which may suggest an instability of the dual field theory.
Similarly, the two-point function \eqref{lift} can be analytically continued to  $\tilde{m} = \pm im > 0$:
\begin{align}
\langle O(k_x,k_y) O(p_x,p_y) \rangle &= \delta(k_x+p_x)\delta(k_y+p_y) k_x^{\tilde{m}} \frac{\Gamma\left(i\frac{k_y^2}{4k_x}+\frac{1}{2} +\frac{\tilde{m}}{2}\right)}{\Gamma\left(i\frac{k_y^2}{4k_x}+\frac{1}{2} -\frac{\tilde{m}}{2}\right)} \label{lift}
\end{align}
and can be directly compared to the one obtained in \cite{Kachru:2008yh}.

We emphasize that unlike the conformal field theories, the two-point functions of the anisotropic scale invariant field theories are not uniquely determined from the symmetry alone, so our computation gives a precise prediction of the two-point functions of the dual anisotropic scale invariant field theory. At the same time, however, any non-minimal coupling of the scalar fields to the background geometry as well as the background matter field would change the form of the correlation functions, so a specification of the coupling is needed to fully determine the two-point functions. Such specification is outside the scope of the bottom up effective field theory approach taken in this paper, but the consistency of the effective field theory approach demand that such corrections should be small compared with the leading order behavior studied in this section.

\section{Discussion}
In this paper, we investigated the possibility of anisotropic scale invariant cosmology within the conventional Einstein gravity. We have shown that it is only possible when the null energy condition is violated. Given a difficulty to break the null energy condition, it would be very important to find an embedding in the ultraviolet completed quantum theories of gravity. Note that the reference \cite{Li:2009pf} suggested that even a less exotic anisotropic Lifshitz geometry is difficult to realize in the string theory.\footnote{A physical reasoning behind the ``no-go theorem" is obscure. Indeed, a smeared (integrated) version of the equations motion do not have any obstruction to obtain Lifshitz type geometry \cite{Nak} (see also \cite{Hartnoll:2009ns}).}

From the dual field theory perspective, in particular for the Euclidean field theories, the unitarity or reflection positivity is not the central dogma. The unitarity is a crucial issue to understand the consistency of the quantum gravity, but for the time-dependent holography like dS/CFT or our anisotropic cosmology, it is yet to be investigated how the unitarity of the bulk theory is encoded in the boundary theory. On the other hand, it is very important to understand how to construct gravity dual of non-unitary field theories from the gravity viewpoint because there are plenty of non-unitary, or non-reflection positive examples of interesting condensed matter systems. Our construction is one approach in this direction.

Finally, it would be interesting to study the realization of anisotropic scale invariant cosmology in the anisotropic gravity. Since the anisotropic scale invariance is already encoded in the action, it may be more suitable to discuss the anisotropic cosmology there than in the conventional Einstein gravity where the violation of the null energy condition is needed.\footnote{A cosmology model with $z=5/3$ based on a conformal Galillean matter coupled with a scalar gravity has been discussed in \cite{Stichel:2009sz}\cite{Stichel:2009mj}.}

\section*{Acknowledgements}
The author would like to thank B.~Freivogel for stimulating discussions on the energy conditions.
The work was supported in part by the National Science Foundation under Grant No.\ PHY05-55662 and the UC Berkeley Center for Theoretical Physics  and World Premier International Research Center Initiative (WPI Initiative), MEXT, Japan.

\appendix

\section{Confluent Hypergeometric functions}
The confluent hypergeometric function $U(a,b;x)$ is a solution of Kummer's equation
\begin{eqnarray}
x\frac{d^2 U}{dx^2} + (b-x) \frac{dU}{dx} - a U = 0 \ 
\end{eqnarray}
with the series expansion
\begin{align}
U(a,b;x) = &x^{1-b} \left[\frac{\Gamma(-1+b)}{\Gamma(a)} + \frac{(-1-a+b)\Gamma(-2+b)}{\Gamma(a)} x + \cdots  \right] \cr
 &+ \frac{\Gamma(1-b)}{\Gamma(1+a-b)} - \frac{a\Gamma(-b)}{\Gamma(1+a-b)} x + \cdots \ .
\end{align}

Alternatively it has an integral representation
\begin{eqnarray}
U(a,b;x) = \frac{1}{\Gamma(a)} \int_0^\infty dt e^{-xt} t^{a-1}(1+t)^{b-a-1}  \ .
\end{eqnarray}
It satisfies 
\begin{eqnarray}
U(a,b;x) = x^{1-b}U(1+a-b,2-b;x) \ .
\end{eqnarray} 

Similarly, the generalized Laguerre function $L(a,b;x)$ satisfies
\begin{align}
x\frac{d^2 L}{dx^2} +(b+1-x)\frac{d L}{dx} + a L = 0 
\end{align}
and has a series expansion:
\begin{align}
\frac{L(a,b;x)}{L(a,b;0)} = _1F_1(-a,b+1;x) = \sum_{n=0}^{\infty}\frac{(-a)_n x^n}{(b+1)_n n!} \ .
\end{align}

\end{document}